\documentstyle[aps,twocolumn]{revtex}

\begin{document}

\input epsf.sty
\newcommand{\infig}[2]{\begin{center}\mbox{ \epsfxsize #1
                       \epsfbox{#2}}\end{center}}
\newcommand{\infigtwo}[3]{\begin{center}\mbox{ \epsfxsize #1
                        \epsfbox{#2} \quad \epsfxsize #1
                        \epsfbox{#3}} \end{center}}
\newcommand{\infigtwon}[4]{\begin{center}\mbox{ \epsfxsize #1
                        \epsfbox{#2} \quad \epsfxsize #3
                        \epsfbox{#4}} \end{center}}

\newcommand{\be}{\begin{equation}}
\newcommand{\ee}{\end{equation}}
\newcommand{\bea}{\begin{eqnarray}}
\newcommand{\eea}{\end{eqnarray}}

\draft

\title{Scaling properties of cavity-enhanced atom cooling}
\author{Peter Horak and Helmut Ritsch}

\address{Institut f{\"u}r Theoretische Physik, Universit{\"a}t Innsbruck, 
Technikerstr.\ 25, A-6020 Innsbruck, Austria.} 
\date{\today}

\maketitle 
 
\begin{abstract} 
We extend an earlier semiclassical model to describe the dissipative motion of
$N$ atoms coupled to $M$ modes inside a coherently driven high-finesse cavity. 
The description includes momentum diffusion via spontaneous emission and cavity
decay. Simple analytical formulas for the steady-state temperature and the
cooling time for a single atom are derived and show surprisingly good agreement
with direct stochastic simulations of the semiclassical equations for $N$ atoms
with properly scaled parameters. A thorough comparison with standard free-space
Doppler cooling is performed and yields a lower temperature and a cooling time
enhancement by a factor of $M$ times the square of the ratio of the atom-field
coupling constant to the cavity decay rate. Finally it is shown that laser
cooling with negligible spontaneous emission should indeed be possible,
especially for relatively light particles in a strongly coupled field
configuration.
\end{abstract}
 

\narrowtext


\section{Introduction}
For a large variety of atoms, laser cooling has proven to be an extremely
successful and widely applicable method to generate fairly large ensembles of
ultra-cold neutral atoms \cite{Review}. 
Nevertheless, some intrinsic effects, like
reabsorption of the spontaneously scattered light, optical pumping to unwanted
states or dipole heating, limit its applicability and the final phase space
density that can be achieved. It has been suggested
\cite{PRL,Vuletic,Lewenstein,doherty97} that these
problems could be partly cured by the help of high-finesse optical resonators to
enhance the cooling process and make it more selective in frequency space as
well as in position space. For single atoms moving in the field of only one or
a few photons in high-$Q$ microscopic resonators these results could
largely be demonstrated experimentally\cite{Rempe,Kimble}. Although this nicely
confirms the theoretical concepts and model approximations, the scaling towards
larger particle and photon numbers as well as large interaction volumes is not
entirely clear. Some initial results on the scaling properties of this cooling
scheme were obtained for ensembles of well trapped particles
\cite{RapidComm}, where
a harmonic approximation for the trapping potential can be assumed. Neglecting
cavity induced momentum diffusion, collective equations for the atomic motion
could be derived from which the scaling properties with atom number and volume
could be read off. Similarly in the very far detuned weak field case, where
spontaneous emission and field induced atom-atom interaction can be neglected,
estimates of the scaling properties could be derived from geometrical
considerations\cite{Vuletic}.

The range of the validity of both of these approaches is, however, not
entirely clear and does not cover a large parameter regime. In many practical
cases neither the tight binding situation allowing a collective description of
the properties of a whole ensemble, nor the weak field assumption treating each
atom separately will be realized. In that case one cannot expect reliable
answers from these models. Only a more general description bridging the gap
from the single atom to the collective description and from a single photon to
a coherent field with many photons should provide adequate answers on the
efficiency and realizability of this cooling scheme.  

In this work we use a quite general semiclassical approach which only relies on
the assumptions of a weak saturation and a not too low temperature  $T\gg
T_{recoil}$, which should be true for most practical setups. The model as
defined in section II consistently includes spontaneous emission, cavity decay,
atom-atom interaction via the field, and realistic mode functions. For a single
atom in a standing wave field simple analytic expressions for the final
temperature and the cooling time are derived in section III. In a rescaled
version these are then applied to the $N$-atom case and tested against
numerical simulations in section IV. In section V we compare the results to
standard Doppler cooling and discuss the possibility of cooling without
spontaneous emission in section VI. Finally, in section VII we briefly discuss
the scaling properties for the case of several degenerate cavity modes.


\section{Semiclassical model}
\label{sec:model}

We consider a dilute gas of $N$ identical two-level atoms of mass $m$, 
transition frequency $\omega_a$, and spontaneous emission rate $\Gamma$
interacting with $M$ modes of a high-finesse optical cavity with resonance
frequencies $\omega_k$, cavity decay rates $\kappa_k$, and mode functions
$f_k(x)$, $k=1..M$, respectively. The mode functions are mutually orthogonal and
fulfill the normalization condition $\int |f_k|^2\, dx=V$, where $V$ is the mode
volume. The spatially averaged single photon Rabi frequency is
denoted by $g$. The cavity is driven by a pump laser of frequency $\omega_p$, 
the spatial overlap of the pump with the mode functions at the driven cavity
mirror yields pump strengths $\eta_k$.

Generalizing previous work \cite{JPB}, we derive a semiclassical model of the
system described by a set of coupled stochastic differential equations (SDEs)
for the atomic positions $x_n$, momenta $p_n$, and mode amplitudes $\alpha_k$,
\begin{mathletters}
\bea
dx_n & = & \frac{p_n}{m}\, dt,\\
dp_n & = & -U_0\Big[{\cal E}(x_n)\nabla_n {\cal E}^*(x_n) 
                    + {\cal E}^*(x_n)\nabla_n {\cal E}(x_n)\Big]\,dt
   \nonumber \\ & &
   + i \gamma\Big[{\cal E}(x_n)\nabla_n {\cal E}^*(x_n) 
                  - {\cal E}^*(x_n)\nabla_n {\cal E}(x_n)\Big]\,dt
   \nonumber \\ & &
   + dP_n, \label{eq1b}\\
 d\alpha_k & = & -\eta_k^* dt 
   + i\Big[ \Delta_k\alpha_k-U_0\sum_n {\cal E}(x_n)f_k^*(x_n)\Big] dt
   \nonumber \\ & &
   - \Big[ \kappa_k\alpha_k+\gamma\sum_n {\cal E}(x_n)f_k^*(x_n)\Big] dt
   + dA_k, \label{eq1c}
\eea
\label{eq1}
\end{mathletters}
where ${\cal E}(x)=\sum_k f_k(x)\alpha_k$ is the field amplitude at position
$x$, $U_0 = \Delta_a g^2/(\Delta_a^2+\Gamma^2)$ the light shift per
photon ($\Delta_a = \omega_p-\omega_a$), 
$\gamma = \Gamma g^2/(\Delta_a^2+\Gamma^2)$ the photon scattering rate,
and $\Delta_k = \omega_p-\omega_k$ the detuning of the $k$th mode from the pump
laser.

The interpretation of the various terms in eqs.~(\ref{eq1}) is rather simple.
The two terms in eq.~(\ref{eq1b}) correspond to the dipole force and the
radiation pressure force, respectively. Equation (\ref{eq1c}) describes the
pumping of the cavity modes, the frequency detuning from the pump laser shifted
according to the coupling to the atoms, and the mode damping due to cavity
decay and photon scattering by the atoms. $dP_n$ and $dA_k$ are white noise
increments which can be characterized by a diffusion matrix  $D_{ij}=\langle
dF_i\, dF_j\rangle$, where $F_i=\{ P_n, A_k\}$. The diffusion matrix exhibits
nontrivial cross correlations between momentum diffusion and cavity amplitude
and phase fluctuations \cite{JPB}. We may decompose the noise terms by writing
\begin{mathletters}
\bea
dP_n & = & dP_n^{\tiny spont} + dP_n^{\tiny ind}, \\
dA_k^r & = & dA_k^{\tiny r,spont} + \sum_n dA_{k,n}^{\tiny r,ind}, \\
dA_k^i & = & dA_k^{\tiny i,spont} + \sum_n dA_{k,n}^{\tiny i,ind}.
\eea
\label{eq2}
\end{mathletters}
The first terms correspond to momentum and mode amplitude fluctuations due to
spontaneous emission of photons into vacuum modes. The second terms describe
the correlated momentum and field fluctuations, which occur, when the $n$th atom
scatters a photon among different cavity modes. The precise form of all noise
terms is given in Appendix \ref{app}.

The set of SDEs (\ref{eq1}) is well suited for a numerical implementation. Note
that the number of equations grows linearly with the number of atoms and modes,
in sharp contrast to full quantum descriptions, which become exceedingly large
for all but the simplest situations. The inclusion of the quantum noise terms,
on the other hand, allows to obtain reliable results, for example, for
steady-state temperatures and trapping times, which cannot be derived from a 
simple classical model. 

Let us now specify the parameter regime we are concentrating on in this paper.
We assume an optical cavity of sufficiently high finesse with large separation
between the various eigenfrequencies. Then we can neglect all modes, which are
not close to resonance with the pump field frequency. For a standing wave
cavity the parameters can be chosen such, that only a single mode will fulfill
this condition. For a ring cavity setup one should include at least the
two degenerate counterpropagating running waves, while in a quasiconfocal or
quasispherical setup many transverse modes can be involved in the dynamics. We
will discuss these situations later in section VII. For
the following, we will specialize the model to a one-dimensional
standing wave setup, neglecting transverse atomic motion.
Therefore, only a single mode function $f(x)=\cos(kx)$ is considered
and we will drop the subindex $k$ in the notation for the field amplitude, 
detuning, and pump strength. Since
the derivation of the SDE is based on an adiabatic elimination of the internal
atomic degrees of freedom, we must ensure a small atomic saturation parameter,
that is, $s = |\alpha|^2 g^2/(\Delta_a^2+\Gamma^2)\ll 1$. To this end we assume
a large atomic detuning $\Delta_a\gg\Gamma$. In order to observe a significant
effect of the atomic motion onto the cavity dynamics, we require $\Delta\approx
-\kappa$ and $N |U_0|\lesssim\kappa$. 
Hence, for large atom numbers $N$, we assume
$|U_0|\ll\kappa$. Although these parameters are known to give rise to cooling
independent of the sign of the atomic detuning (and hence the sign of $U_0$),
we will in the following use negative values of $U_0$ since, in a full
model, this additionally gives rise to three dimensional trapping.

\section{Analytical results for a single atom}
\label{sec:single}

Before we discuss the results obtained from numerical simulations of the SDEs,
we will briefly summarize the analytical results obtained earlier for a
single atom in a single mode of a standing wave cavity and specialize these to
the present parameters.
As we will see later, this provides an excellent basis for the discussion of the
multi-atom case.

The analytical results \cite{PRL,PRA} were derived under the assumption of weak
pumping $\eta$ such that there is at most one excitation in the system.
However, it can be shown that under the conditions required for the adiabatic
elimination of the atomic excited state, these results also hold for higher
photon numbers as long as $s\ll 1$. Thus, we may apply the formulas in the
semiclassical limit discussed here. For atomic velocities $v$ fulfilling
$kv<\kappa$ the force on a moving atom can be expanded to lowest order in
$kv/\kappa$ as
\be
F = F_0 + F_1 v,
\ee
where $F_0$ is the stationary dipole force from eq.~(\ref{eq1b}) for a fixed
atomic position, and $F_1$ is the linear friction coefficient,
\be
F_1(x) = -2 k^2 \cos(k x)^2\sin(k x)^2 \frac{\eta^2 U_0^2}{\kappa^4}.
\label{eq:fricx}
\ee
Neglecting localization effects, the position-averaged friction force thus reads
\be
\overline{F_1} = -k^2\frac{\eta^2 U_0^2}{4\kappa^4}.
\label{eq:fric1}
\ee
Similarly, the averaged momentum diffusion coefficient is found to be
\be
\overline{D} = k^2\kappa\frac{\eta^2 U_0^2}{8\kappa^4}
\ee
and hence the steady-state temperature is
\be
k_B T = -\frac{\overline{D}}{\overline{F_1}} = \frac{\kappa}{2}.
\label{eq:temp}
\ee
Thus, we can get sub-Doppler cooling for $\kappa<\Gamma$.

Finally, from the friction coefficient we obtain an estimate for the cooling
time $\tau_c$ defined from the exponential decrease of the kinetic energy $E(t)$
of the form
\be
E(t) = \left[E(0)-\frac{k_B T}{2}\right] \exp(-t/\tau_c) + \frac{k_B T}{2}
\label{eq:exp}
\ee
with
\be
\tau_c = \frac{m}{2|\overline{F_1}|} = \frac{\kappa^4}{\eta^2 U_0^2} 
   \omega_R^{-1}.
\label{eq:Tc}
\ee
Here $\omega_R = \hbar k^2/(2m)$ is the recoil frequency. Hence we get faster
cooling for larger cavity field intensity and larger optical potentials per
photon. Moreover, the cooling time is proportional to the atomic mass such that
the scheme works best for relatively light atoms.


\section{Cooling of many atoms}

Let us now consider $N$ atoms. As they all couple to the same mode, one might
expect that in general they will perturb each other and interact via this mode
\cite{Pellizzari,Hemmerich,Parkins,Fischer}.
In fact, for a few atoms and a few photons inside a high-finesse cavity such
effects have recently been demonstrated experimentally \cite{RempeMult}.
In order to investigate the cooling properties of many atoms, we numerically
integrate the SDEs~(\ref{eq1}) and average over many different realizations.
Choosing a flat initial distribution of atoms in space and a large initial
kinetic energy we calculate the time evolution of the kinetic energy. In the
following we will concentrate on the steady-state temperature and, in
particular, on the cooling time scale $\tau_c$.

\begin{figure}
\infig{6.5cm}{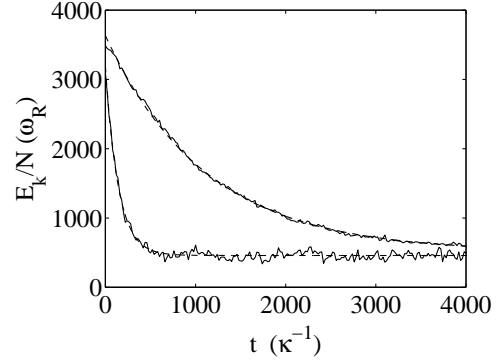}
\caption{Time evolution of the mean kinetic energy per atom obtained from
numerical simulations of the SDEs averaged over 100 realizations (solid curves).
The dashed curves are exponential fits. Lower curves: $N=1$, upper curves:
$N=10$. The parameters are
$NU_0=-0.6\kappa$, $N^2\gamma=0.03\kappa$, $\Delta=-0.6\kappa$,
$\eta=3\sqrt{N}\kappa$, $\kappa=415\omega_R$.} 
\label{figA}
\end{figure}

In Fig.~\ref{figA} we plot the numerically obtained mean kinetic energy as a
function of time for a single atom and for ten atoms. For the many atom case we
rescale the parameters such that $NU_0$ and the optical potential depth 
(proportional $U_0\eta^2/\kappa^2$) are constant. We find that the curves are
very well approximated by an exponential fit as in eq.~(\ref{eq:exp}) with 
cooling times $\tau_c=142\kappa^{-1}$ ($N=1$) and $\tau_c=1110\kappa^{-1}$ 
($N=10$). For the given parameters, the analytical estimate given by
eq.~(\ref{eq:Tc}) yields $\tau_c=118.6\kappa^{-1}$ and $1186\kappa^{-1}$,
respectively, which is surprisingly good. Therefore, the single atom cooling
time appears to make sense for the many atom case. However, the operating
conditions differ for varying atom numbers, according to the condition that
$NU_0$ remains constant.

\begin{figure}
\infig{6.5cm}{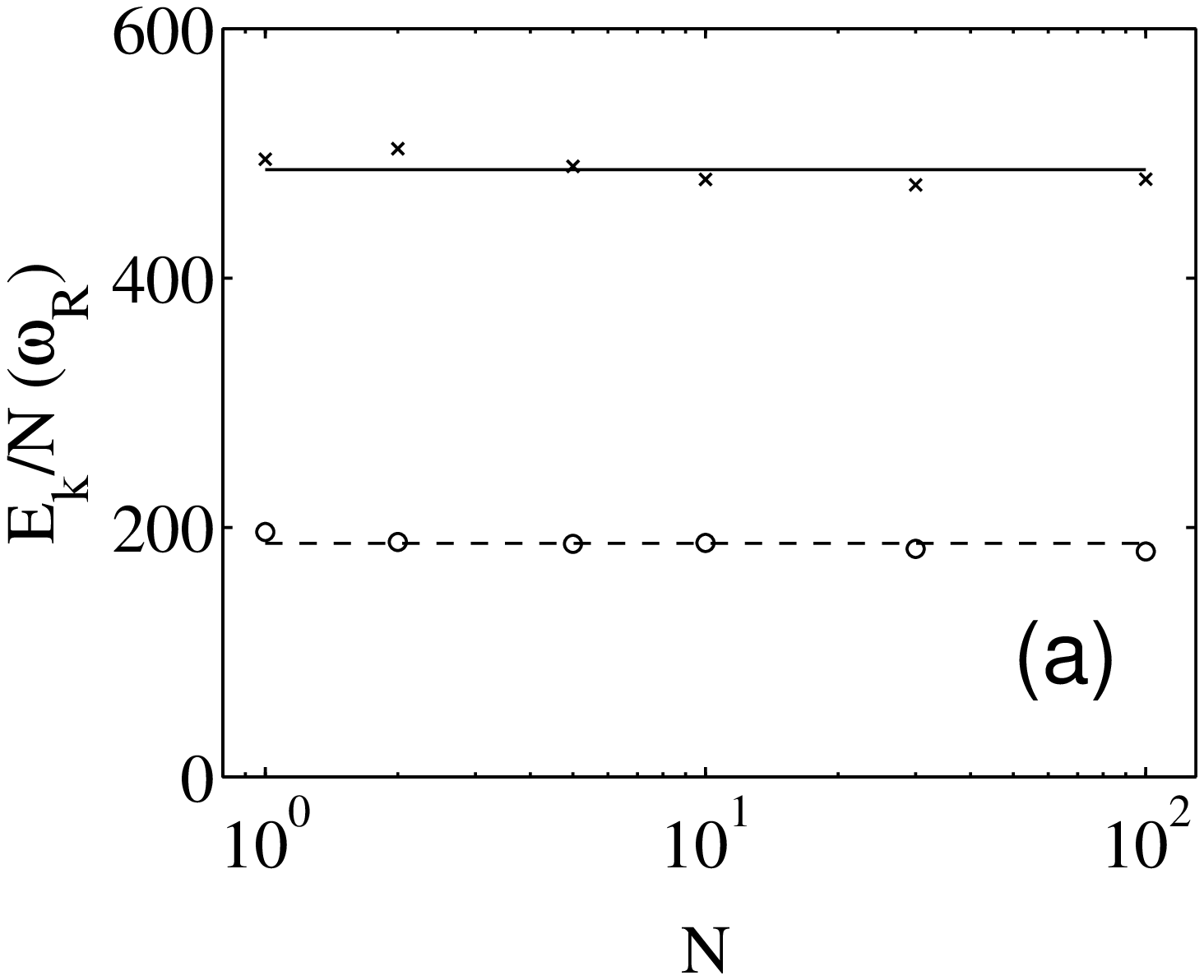}
\infig{6.5cm}{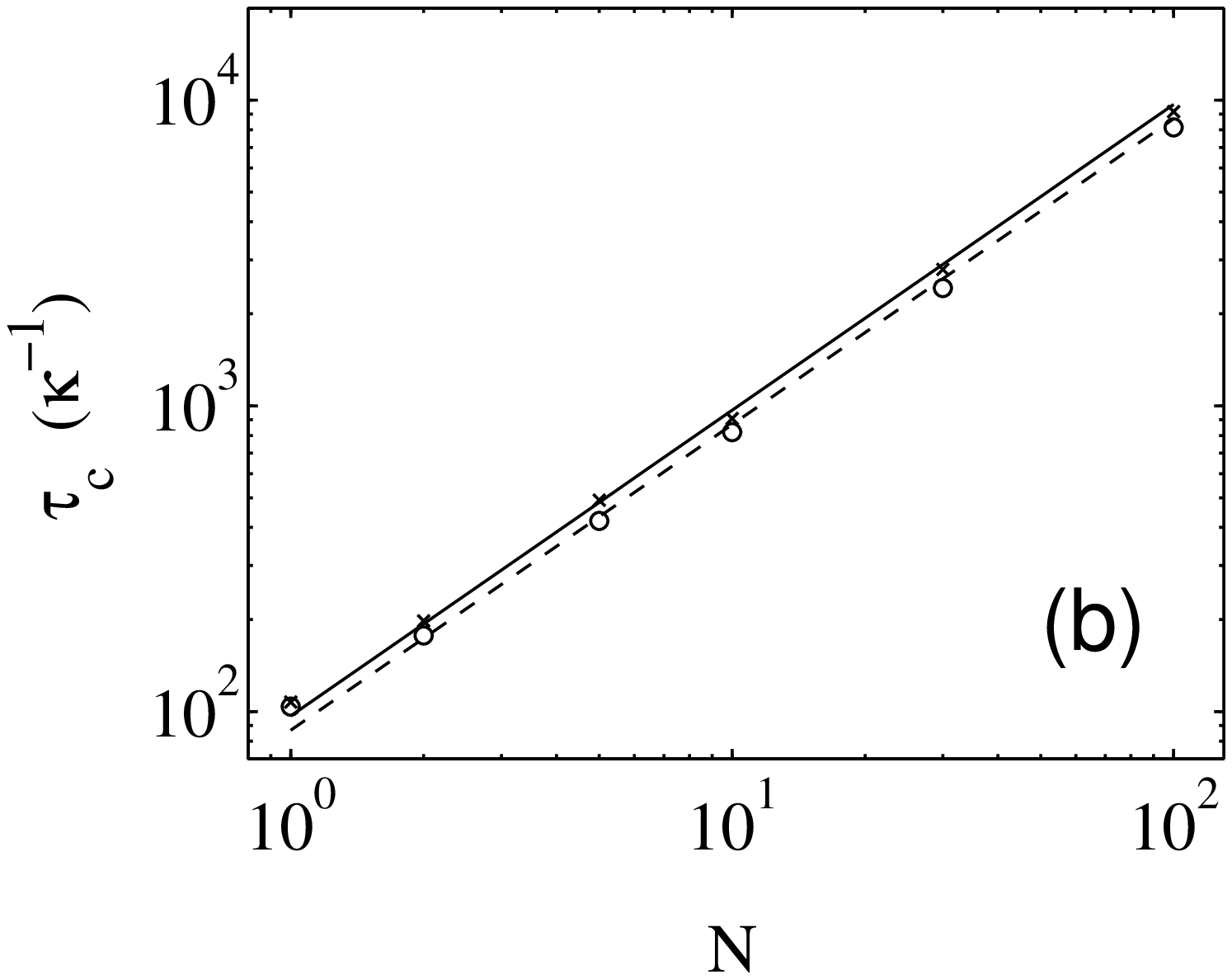}
\caption{Steady-state kinetic energies per atom (a) and cooling times $\tau_c$ 
(b) for atom numbers from 1 to 100. 
Crosses show the numerical results for $\Delta=-0.6\kappa$, circles for 
$\Delta=-\kappa$. All other parameters are the same as 
in Fig.~\protect\ref{figA}. The solid and dashed lines are linear fits to the
numerical data.}
\label{figB}
\end{figure}

The steady-state kinetic energies obtained from the exponential fits of
Fig.~\ref{figA} are $466\omega_R$ ($N=1$) and $510\omega_R$ ($N=10$). While
these are approximately of the same size, they are significantly larger than
the value of $\kappa/4\approx 100\omega_R$ obtained from eq.~(\ref{eq:temp}).
However, eq.~(\ref{eq:temp}) was derived assuming $U_0\ll\kappa$ and a flat
spatial distribution in the steady state. Both of these conditions are violated
for the present parameters.

It should also be emphasized that an exponential fit as in Fig.~\ref{figA} only
works for sufficiently hot atoms where the friction force (\ref{eq:fricx}) can
be replaced by its mean value and can therefore be treated as constant. For
very cold and well localized atoms the position dependence of the friction must
be taken into account. In fact, eq.~(\ref{eq:fricx}) predicts that the friction
force vanishes completely at the bottom of the potential wells. Hence, if in
the course of the cooling process the localization increases, the time scale of
the cooling increases and the kinetic energy no longer decreases according to
an exponential law. For such parameters, a unique cooling time $\tau_c$ can
only be deduced from the initial stages of the cooling while the atoms are
still uniformly distributed in space.

For the same scaling of the parameters with $N$ as discussed above, we plot the
steady-state kinetic energy and the cooling time for particle numbers from one
to 100 in Fig.~\ref{figB}. We find an approximately constant final temperature
and a cooling time proportional to $N$. Taking the scaling of $U_0$ with the
particle number $N$ into account, these results agree with the predictions of
the single-atom estimates of the previous section.

\begin{figure}
\infig{6.5cm}{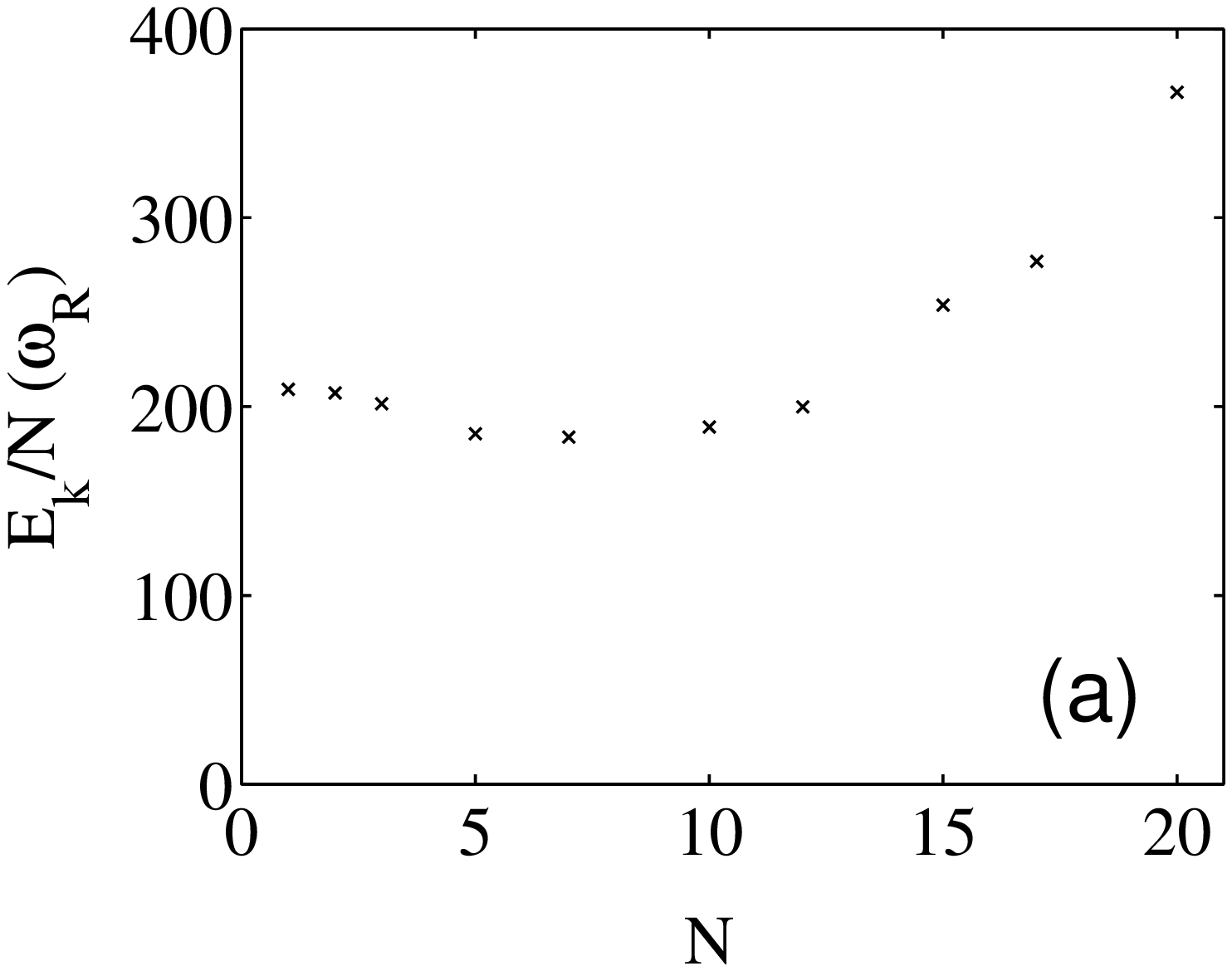}
\infig{6.5cm}{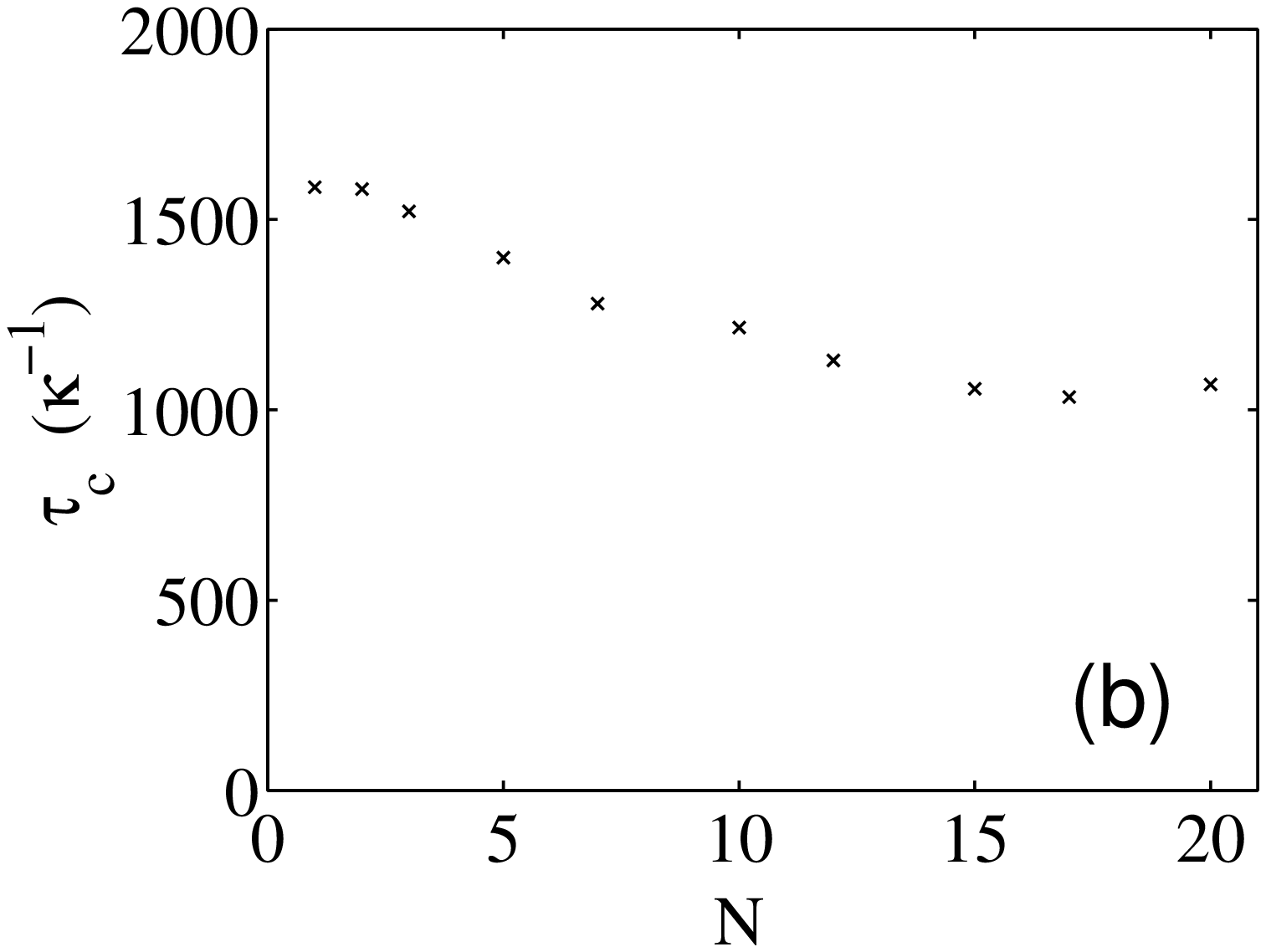}
\caption{Steady-state kinetic energies per atom (a) and cooling times $\tau_c$ 
(b) for atom numbers from 1 to 10. Parameters are $U_0=-0.05\kappa$, 
$\gamma=2.5\,10^{-4}\kappa$, $\Delta=-\kappa$,
$\eta=10\kappa$, $\kappa=415\omega_R$.}
\label{figC}
\end{figure}

In Fig.~\ref{figC}, a different scaling of the system parameters is applied. We
choose a small value of $U_0$ and keep all parameters constant while changing
the atom number. For the steady-state temperature we obtain approximately
constant values as long as $N|U_0|\lesssim \kappa/2$ ($N\lesssim 10$). 
For $10\lesssim N \lesssim 25$ the
frequency shift induced by the atoms is of the order of the cavity detuning
$\Delta=-\kappa$, which reduces the cooling efficiency and consequently
increases the steady-state temperature. For even larger atom numbers the cavity
is shifted into positive detuning where the friction force changes sign and the
atoms are accelerated. In this limit no stationary state is achieved and no
cooling time can be defined.

For the same parameters, figure~\ref{figC}(b) shows the cooling time for varying
atom number. We see that $\tau_c$ slightly decreases with
increasing $N$. Close inspection of the numerical data shows, however, that this
decrease can be entirely explained by an increase of the cavity photon number
according to the increase of $NU_0$, which shifts the cavity closer to resonance
for larger atom numbers. Keeping the photon number constant, e.g., by adjusting
the pump strength $\eta$, would thus give rise to a cooling time essentially
independent of $N$.

Therefore, the numerical simulations show that for $NU_0<\kappa$ the
cooling time for an ensemble of $N$ atoms is of the same order of magnitude as
the cooling time for a single atom. This suggests that the individual atoms in
the cloud are cooled independently from each other, although they are all coupled
to the same cavity mode.

For the case of $NU_0\ll\kappa$ (no effect of the atoms on the cavity mode)
this has
also been predicted by Vuletic and Chu \cite{Vuletic}. The basic argument there
is that the cavity enhances backscattering of photons by moving atoms according
to the Doppler shift, similar to free-space Doppler cooling. We now see that
the cooling of atoms is nearly independent of the atom number also in cases
where the cavity photon number is significantly changed by the presence of the
atoms  ($NU_0\lesssim\kappa$). We attribute this to the fact that, considering
the motion of an individual atom out of the whole ensemble, only the change in
the cavity field induced by that atom itself is correlated with the atomic
motion, while the influence of all other atoms is uncorrelated and hence on
average cancels. Note that this argument is based on the assumption of no
correlation between the motion of different atoms. Thus, for deep (harmonic)
optical potentials in the classical limit where all particles oscillate with
the same frequency a completely different scaling with the atom number has been
found \cite{RapidComm}. For the parameters considered here, the anharmonicity
of the sinusoidal potential implies a different oscillation frequency for
different atoms. Together with the independent part of the momentum diffusion
this ensures fast dephasing between the individual atomic oscillations. Despite
the fact that they commonly interact with the same field these mechanisms
guarantee that the atoms  can be considered as independent. Such decorrelation
(phase space mixing) has also been shown to be important for stochastic cooling
of atoms in optical traps\cite{Raizen,Davidson}, which has been suggested as
another alternative to cool atoms without spontaneous emission.


\section{Comparison with free-space Doppler cooling}

Since we have seen that the single atom formulas of Sec.~\ref{sec:single}
provide a good approximation for the many atom case, too, we will investigate
these in the following a bit further.

Let us first introduce the spatially averaged atomic saturation $s$ as
\be
s = |\alpha|^2\frac{g^2 \overline{\cos(kx)^2}}{\Delta_a^2+\Gamma^2}
\approx \frac{\eta^2U_0^2}{4\kappa^2 g^2}.
\ee
For the validity of the SDEs (\ref{eq1}) we required $s\ll 1$. We can then
rewrite the averaged friction force (\ref{eq:fric1}) in the form
\be
\overline{F_1} = -s k^2\left(\frac{g}{\kappa}\right)^2
\label{eq:fric2}
\ee
and the cooling time as
\be
\tau_c = \frac{\kappa^2}{4sg^2}\omega_R^{-1}.
\label{eq:tc2}
\ee
Hence, for a given atom and a given saturation the maximum friction force
and minimum cooling time $\tau_c$ is limited by the ratio of the atom-cavity
coupling $g$ and the cavity decay rate $\kappa$. Cooling is most efficient if
this ratio is large, that is, for high-finesse cavities. 

For standard Fabry-Perot resonators \cite{Yariv} with mirror curvature larger
than the cavity
length $l$, the mode waist scales as $l^{1/4}$ and thus the mode volume $V$
as $l^{3/2}$. Since $g\propto 1/\sqrt{V}$ and $\kappa\propto 1/l$ we find
$(g/\kappa)^2\propto \sqrt{l}$. Thus, cavity enhanced cooling seems more
favorable for longer cavities, which also would allow to include more atoms.
However, this route is limited by the capture range of the cooling given by
$kv<\kappa$. Thus, for a small cavity decay rate only already very cold atoms
are subject to the cavity cooling force.

Let us now compare eqs.~(\ref{eq:fric2}) and (\ref{eq:tc2}) with the
corresponding results for free-space Doppler cooling \cite{CCT}. This provides
optimum cooling for $\Delta_a=-\Gamma$. The friction force is given by
\be
\overline{F_1^{Dopp}} = -s k^2,
\ee
and the cooling time is therefore
\be
\tau_c^{Dopp} = \frac{1}{4s}\omega_R^{-1}.
\ee
We note that the only difference with the cavity-induced cooling is the absence
of the factor $(g/\kappa)^2$. Hence, only in high-Q cavities with $g>\kappa$ the
cooling scheme discussed here is faster than standard Doppler cooling. The
achievable steady-state temperatures of the order of $\kappa/2$, on the other
hand, can be significantly lower than the Doppler limit of $\Gamma/2$ even for
cavities with $g<\kappa$ but at the expense of a smaller capture range as
already discussed before.


\section{Spontaneous atomic decays}

Cavity-induced cooling has been suggested as a possible means to cool particles
without closed optical transitions such as certain atomic species or molecules.
The motivation behind this proposal is that the cooling scheme relies on induced
internal transitions of the particle rather than on spontaneous emission of
photons. On the other hand, equation~(\ref{eq:fric2}) shows that for large
atomic detuning and correspondingly small saturation, which is required for a
small photon scattering rate $\gamma$, the cooling time also becomes very long.
The figure of merit in this respect is the number of photons $N_{ph}$ which are
spontaneously scattered per particle during one cooling time $\tau_c$. Only for
$N_{ph}\approx 1$ effective cooling of, for example, a molecule becomes
feasible.

Using the results of the previous section, we can easily obtain an estimate for
this number from the definition $N_{ph}=2\Gamma s \tau_c$ yielding
\be
N_{ph}= \frac{\gamma\kappa^2}{2\omega_R U_0^2}
      =\frac{1}{2}\frac{\Gamma}{\omega_R}\left(\frac{\kappa}{g}\right)^2.
\ee
Note that this result is independent of the pump strength $\eta$ and the atom
detuning $\Delta_a$, and is entirely fixed by atom and cavity properties.

As in the previous section, it is interesting to compare this with the
corresponding result for standard Doppler cooling given by
\be
N_{ph}^{Dopp}=\frac{1}{2}\frac{\Gamma}{\omega_R} \gg 1.
\ee
Again, the only difference is the missing factor of $(\kappa/g)^2$.
Therefore, only for $g>\kappa$ the number of spontaneously emitted photons per
cooling time is reduced with respect to free-space Doppler cooling. Despite the
fact that cavity-induced cooling works at arbitrarily large atomic detuning, the
increased cooling time counteracts this reduction of the scattering rate
$\gamma$, such that the overall efficiency falls behind Doppler cooling for
cavities of not sufficiently high finesse.
For high-Q cavities, on the other hand, the reabsorption problem, which limits
the achievable particle density in free-space laser cooling, is strongly
reduced too, as a photon which has decayed through the cavity
mirror cannot be reabsorbed \cite{Kim}. 


\section{Multi-mode cavities}

Let us now return to the more general situation where the optical cavity
supports several (nearly) degenerate modes as described by the SDEs~(\ref{eq1}).

In a first step, we use the friction coefficient obtained for a single atom in
a ring cavity \cite{ringPRA} to derive analytical estimates analogously to
those in sections III and VI for a single-mode standing-wave cavity. We
find\footnote{Note that for consistency with the notations used here, the mode
functions and the pump strengths used in Ref.~\protect\cite{ringPRA} have to be
multiplied by a factor of $1/\sqrt{2}$.}
\bea
s & = & \frac{\eta^2 U_0^2}{4\kappa^2 g^2},\\
k_B T & = & \frac{\kappa}{2}, \\
\tau_c & = & \frac{1}{2} \frac{\kappa^4}{\eta^2 U_0^2} \omega_R^{-1}, \\
N_{ph} & = & \frac{1}{4}\frac{\Gamma}{\omega_R}\left(\frac{\kappa}{g}\right)^2
\eea
Hence, for the same mean atomic saturation we obtain the same steady-state
temperature but only half the cooling time and only half the number of
spontaneously scattered photons per atom compared to the single-mode cavity.

This suggests that using $M$ degenerate optical modes, for instance in a
near-confocal or near-planar setup, improves the efficiency of
cavity-enhanced cooling by a factor of $M$. In fact, numerical simulations of
ten atoms and one to ten modes have confirmed this simple scaling. Therefore, 
a multi-mode cavity reduces the cooling time and the number of incoherently
scattered photons per atoms by a factor of $M(g/\kappa)^2$ compared to
free-space Doppler cooling.
Finally, it should be emphasized that in this case and
using again $\Delta\approx -\kappa$, cooling occurs if $MN|U_0|$ (instead
of $N|U_0|$ for a single mode) is of the order of or smaller than $\kappa$.


\section{Conclusions}

In this paper we have investigated the possibility of cooling a cloud of neutral
particles by a cavity-enhanced cooling scheme. Numerical simulations of the
stochastic differential equations describing the situation of many two-level
atoms confined in a single-mode standing-wave optical resonator show that the
friction force and therefore the cooling time is of the same order of magnitude
as for a single atom for the same system parameters. However, the operating
regime which is ideal for one atom cannot be used for many atoms, so that the
best achievable cooling time increases approximately linearly with the atom
number. The analytical results
obtained earlier for that situation can thus be specialized to the parameter
regime relevant for many-particle cooling and have been shown to agree well with
the numerical results. Comparison with the analogous expressions for free-space
Doppler cooling indicate that cavity-induced cooling is mainly favorable for
high-finesse optical cavities fulfilling $g>\kappa$. Numerical simulations also
suggest that the cooling efficiency is increased in cavities which support many
degenerate optical modes.

Using the parameters of the high-Q resonators used in recent cavity-QED
experiments \cite{Rempe,Kimble} one obtains spontaneously scattered photon
numbers per cooling time as low as 5-10. Replacing the heavy rubidium or cesium
atoms by lithium, for example, therefore yields photon numbers below one which
would allow efficient cooling without spontaneous emission. Similarly, only the
lightest molecules form possible candidates for this cooling scheme, with the
further difficulty of the reduced molecular dipole moment between any given
ground and excited state. In any case, according to the small size of these
resonators only very few particles can be trapped and cooled simultaneously.

With larger cavities the efficiency of cavity-induced cooling falls behind that
of Doppler cooling. However, these systems are well suited to demonstrate the
cooling effect on large ensembles by itself. For example, assuming sodium
atoms, an atomic saturation of $s=0.1$, a cavity decay rate $\kappa$ of the
order of one MHz and a coupling constant $g$ of about 100 kHz\cite{Grangier},
cooling times of the order of a ms should be achievable.

An alternative way to reach small mode volumes combined with large finesse is
the use of evanescent wave fields generated by total internal reflection inside
microoptical devices. Examples range from optical microspheres,  thin high
index surface layers or optical band gap guides on surfaces (integrated
optics). Such devices look rather promising for the realization of dissipative
walls or cooling schemes in microtraps.


\acknowledgments

We thank R.\ Grimm
for stimulating discussions. This work was supported by the Austrian Science
Foundation FWF under projects P13435 and SFB ``Control and Measurement of
Coherent Quantum Systems''.


\appendix

\section{Correlated momentum and cavity fluctuations}
\label{app}

In the following we give the full expressions for the correlated noise terms
introduced in Sec.~\ref{sec:model}. The fluctuations induced by spontaneous
emissions read
\bea
dP_n^{\tiny spont} & = & k_0\sqrt{2\gamma |{\cal E}(x_n)|^2}\, dW_n, \\
dA_k^{\tiny r,spont} & = & \sqrt{\kappa/2}\, dW_k^r,\\
dA_k^{\tiny i,spont} & = & \sqrt{\kappa/2}\, dW_k^i,
\eea
where $k_0=\omega_p/c$. The noise terms according to rescattering of photons
within the cavity modes are given by
\bea
\left(
   \begin{array}{c}
      dP_n^{\tiny ind}\\
      dA_{k,n}^{\tiny r,ind}\\
      dA_{k,n}^{\tiny i,ind}
   \end{array}
\right) & = & 
\sqrt{2\gamma}\, {\rm Re}\left\{v_n e^{i\phi_n/2}\right\}\,dW_n^+ \nonumber\\
& & 
+\sqrt{2\gamma}\, {\rm Im}\left\{v_n e^{i\phi_n/2}\right\}\,dW_n^-,
\eea
where $v_n$ and $\phi_n$ are defined by
\be
v_n = 
\left(
   \begin{array}{c}
      \nabla_n{\cal E}(x_n)\\
      -i f_k(x_n)/2\\
      f_k(x_n)/2
   \end{array}
\right),
\quad\quad
e^{i\phi_n} = \frac{\nabla_n{\cal E}^*(x_n)^2}{|\nabla_n{\cal E}(x_n)|^2}.
\ee
Here $dW_n$ is a three dimensional, all other $dW_x^y$ are one-dimensional real
Gaussian stochastic variables of mean zero and variance one.


\end{document}